\documentclass[epj]{webofc}
\usepackage[varg]{txfonts}   
\usepackage{pifont}
\usepackage{xspace}
\def \p{\phantom{2}}\def \px{\p}\def \py{\px\px}

\newcommand{\gv}{\mbox{GeV}}

\newcommand{\ppm}{\pi^+ \pi^-}

\newcommand{\mbo}[1]{$#1$ }

\newcommand{\power}[1]{\times 10^{#1} }

\newcommand{\amu}{a_\mu }
\newcommand{\amuh}{a_\mu^{\rm had} }
\newcommand{\bea}{\begin{eqnarray}}
\newcommand{\eea}{\end{eqnarray}}
\newcommand{\epo}{\;. }

\newcommand{\FFa}{{\cal F}_{\pi^0\gamma^*\gamma^*}}

\newcommand{\FFac}{{\cal F}_{\pi^0\gamma^*\gamma}}
\newcommand{\FFab}{{\cal F}_{\pi^0\gamma\gamma^*}}
\newcommand{\FFabc}{{\cal F}_{\pi^0\gamma\gamma}}

\newcommand{\bary}{\begin{array}}
\newcommand{\eary}{\end{array}}

\newcommand{\amuexp}{a_\mu^{\mathrm{exp}}}
\newcommand{\amuthe}{a_\mu^{\mathrm{the}}}

\woctitle{MESON2018 - the 15$^\textrm{th}$ International Workshop on Meson Physics}
\begin{document}
\selectlanguage{english}
\title{The role of mesons in muon $g-2$}%
%
%

\author{Fred Jegerlehner\inst{1,2}\fnsep\thanks{\email{fjeger@physik.hu-berlin.de}} 
}

\institute{
Deutsches  Elektronen--Synchrotron (DESY), Platanenallee 6, D--15738 Zeuthen, Germany
\and
Humboldt--Universit\"at zu Berlin, Institut f\"ur Physik, Newtonstrasse 15, D--12489 Berlin,
Germany
          }

\abstract{%
The muon anomaly $a_\mu=(g_\mu-2)/2$ showing a persisting
3 to 4 $\sigma$ deviation between the SM prediction and the experiment
is one of the most promising signals for physics beyond the SM.  As is
well known, the hadronic uncertainties are limiting the accuracy of
the Standard Model prediction. Therefore a big effort is going on to
improve the evaluations of hadronic effects in order to keep up with
the 4-fold improved precision expected from the new Fermilab
measurement in the near future. A novel complementary type experiment
planned at J-PARC in Japan, operating with ultra cold muons, is
expected to be able to achieve the same accuracy but with completely
different systematics. So exciting times in searching for New Physics
are under way. I discuss the role of meson physics in calculations of
the hadronic part of the muon g-2.  The improvement is expected to
substantiate the present deviation $\Delta a_\mu^{\rm New \
Physics}=\Delta a_\mu^{\rm Experiment}- \Delta a_\mu^{\rm Standard \
Model}$ to a 6 to 10 standard deviation effect, provided hadronic
uncertainties can be reduce by a factor two. This concerns the
hadronic vacuum polarization as well as the hadronic light-by-light
scattering contributions, both to a large extent determined by the low
lying meson spectrum. Better meson production data and progress in
modeling meson form factors could greatly help to improve the
precision and reliability of the SM prediction of $a_\mu$ and thereby
provide more information on what is missing in the SM. 
}
\thispagestyle{empty}
\begin{flushright}
\large
DESY~18-152,~~HU-EP-18/27\\
September 2018
\end{flushright}

\vfill

\begin{center}
{\Large\bf
The Role of Mesons in Muon $g-2$}\\[1.5cm]
{Fred Jegerlehner}\\[1cm]
{Deutsches  Elektronen--Synchrotron (DESY), Platanenallee 6,\\ D--15738 Zeuthen, Germany\\
Humboldt--Universit\"at zu Berlin, Institut f\"ur Physik, Newtonstrasse 15,\\ D--12489 Berlin,
Germany}

\vfill

{\bf Abstract}\\[3mm]

\begin{minipage}{0.8\textwidth}
The muon anomaly $a_\mu=(g_\mu-2)/2$ showing a persisting
3 to 4 $\sigma$ deviation between the SM prediction and the experiment
is one of the most promising signals for physics beyond the SM.  As is
well known, the hadronic uncertainties are limiting the accuracy of
the Standard Model prediction. Therefore a big effort is going on to
improve the evaluations of hadronic effects in order to keep up with
the 4-fold improved precision expected from the new Fermilab
measurement in the near future. A novel complementary type experiment
planned at J-PARC in Japan, operating with ultra cold muons, is
expected to be able to achieve the same accuracy but with completely
different systematics. So exciting times in searching for New Physics
are under way. I discuss the role of meson physics in calculations of
the hadronic part of the muon g-2.  The improvement is expected to
substantiate the present deviation $\Delta a_\mu^{\rm New \
Physics}=\Delta a_\mu^{\rm Experiment}- \Delta a_\mu^{\rm Standard \
Model}$ to a 6 to 10 standard deviation effect, provided hadronic
uncertainties can be reduce by a factor two. This concerns the
hadronic vacuum polarization as well as the hadronic light-by-light
scattering contributions, both to a large extent determined by the low
lying meson spectrum. Better meson production data and progress in
modeling meson form factors could greatly help to improve the
precision and reliability of the SM prediction of $a_\mu$ and thereby
provide more information on what is missing in the SM. 
\end{minipage}
\end{center}
\vfill
\noindent\rule{8cm}{0.5pt}\\
$^*$ Invited talk  MESON2018 - 15th International Workshop on Meson Physics,
7-12 June 2018, Krak\'ow, Poland.

\setcounter{page}{0}
\def\epjrunnhead{\markboth{submitted to EPJ Web of Conferences}{FCCP2015 -
 Workshop on ``Flavour changing and conserving processes'' 2015}}%
\let\ProcessRunnHead=\epjrunnhead
\newpage
%
\maketitle
%
%
\section{Introduction}
The anomalous magnetic moment (AMM) of the muon $a_\mu=(g_\mu-2)/2$
is one oft the most precisely measured quantities in
particle physics. A very precise measurement~\cite{Bennett:2006fi}
confronts a very precise prediction, revealing a
3 to 4 \mbo{\sigma} discrepancy of the Standard Model (SM) value. It is pure loop
physics, testing virtual quantum fluctuations in depth. New
experiments~\cite{Grange:2015fou,Mibe:2011zz} expected to reach 140
ppb accuracy likely will enhance the significance of the deviation
substantially.

At the present/future level of precision \mbo{a_\mu} depends on all
physics incorporated in the SM: electromagnetic, weak, and strong
interaction effects and beyond that on \textit{all possible new physics}
we are hunting for. For an illustration see e.g. Figs.~13 and 14 in~\cite{Jegerlehner:2018zrj},
which compare physics sensitivities for the muon and the electron, and
unveil the much higher sensitivity of $a_\mu$ on effects beyond QED.

\begin{figure}[h]
\centering
\includegraphics[height=2.4cm]{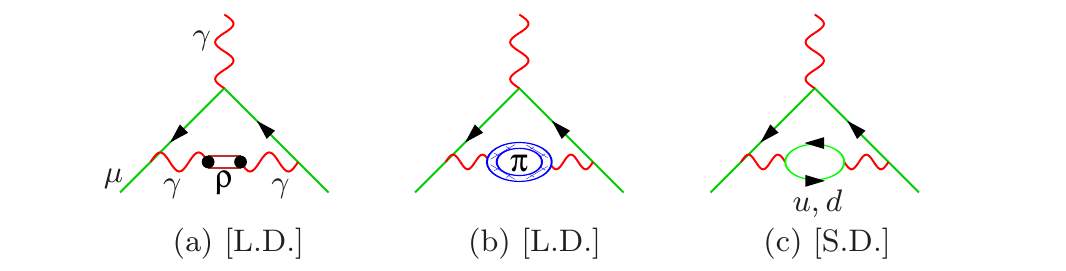}
\caption{\small Leading is the hadronic photon vacuum
polarization of \mbo{O(\alpha^2)}. Diagrams a) and b) show
possible  effective model contributions, VMD and sQED, respectively,
and diagram c) the pQCD tail. The safe method of its evaluation
is a dispersion relation in conjunction with experimental $e^+e^- \to
\gamma^* \to \mathrm{hadrons}$ data or lattice QCD (in progress).}
\label{fig-01}
\end{figure}

\begin{figure}[h]
\centering
\includegraphics[height=2.4cm]{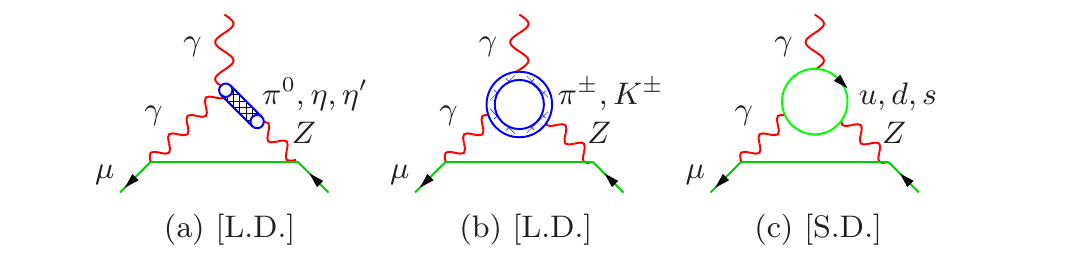}
 \caption{\small Mixed weak hadronic effects. Again we have low energy
effective theory diagrams (L.D.) and the quark-loop diagram
(S.D.). Only the VVA vertex beneath the $\pi^0 Z$
coupling contributes since VVV$\equiv$0. As manifest on the level of the quarks, anomaly
cancellation is at work, which implies that potentially large effects
\mbo{O(\alpha\,G_\mu\,m_\mu^2 \ln M^2_Z/m^2_\mu)} cancel. Therefore
small and well under control.}
\label{fig-02}
\end{figure}

\begin{figure}[h]
\centering
\includegraphics[height=2.4cm]{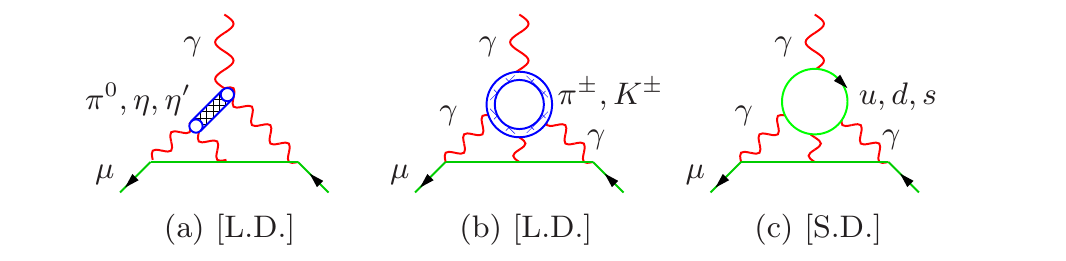}
\caption{\small Hadronic light--by--light scattering of \mbo{O(\alpha^3)}. 
Diagrams (a) and (b) represent the long distance (L.D.) contributions,
diagram (c) involving a quark loop which yields the short distance
(S.D.) tail. Internal photon lines are dressed by $\rho-\gamma$
mixing. This is the most challenging part and also suffers from conceptual problems.}
\label{fig-03}
\end{figure}

The precision of the SM prediction is limited by
substantial hadronic photon vacuum polarization (HVP), Fig.~\ref{fig-01},
while hadronic electroweak (HEW) effects, Fig.~\ref{fig-02},
are small and well mastered. The most difficult and challenging are the hadronic
light-by-light (HLbL) contributions, Fig.~\ref{fig-03}.
Figures~\ref{fig-01},\ref{fig-02} and \ref{fig-03} illustrate the
need for hadronic effective modeling of the dominant long distance
(L.D.) piece, while the short distance (S.D.) tail is calculable by
perturbative QCD (pQCD) [quark-loops], in principle.  The AMM
of the muon is a hot topic these days in view of
two new muon $g-2$ experiments to come. A muon spinning in a
homogeneous magnetic field $\vec{B}$ in absence of an electric field
$\vec{E}$ shows a Larmor spin precession frequency
\mbo{\vec{\omega}} directly proportional to \mbo{\vec{B}}:
$\vec{\omega}_a= \frac{e}{m}\,[a_\mu \vec{B} ]\epo $ The new
Fermilab experiment is improving the ``magic energy'' technique, based
on tuning the beam energy to nullify the electric focusing field
$\vec{E}$ coefficient $(a_\mu-1/(\gamma^2-1))=0$
($\gamma$ the Lorentz factor), and the planned J-PARC experiment
attempting to work in a strict $\vec{E}=0$ environment. The first method
requires ultra relativistic muons (CERN, BNL, Fermilab)), the second
novel concept will work with ultra cold muons (J-PARC) and has very
different systematics.

Then the AMM measurement amounts to measuring the Larmor precession
frequency of the circulating muons and the magnetic field by the
nuclear magnetic resonance method (Larmor precession of protons in a
H$^2$O sample) and the present precision is expected to improve by a
factor 4. The present mismatch $\Delta \amu=\amuthe-\amuexp=(-30.6\pm
7.6)\power{-10}$ would increase to a $6\, \sigma$ deficiency of the SM
prediction if theory is taken as today and the central value would not
move.  Improving theory by reducing the hadronic
uncertainty by a factor 2 could result in a significance of
$11\,\sigma$.

A general introduction I have presented recently
in~\cite{Jegerlehner:2018zrj} (see also my recently actualized
book~\cite{Jegerlehner:2017gek}) and the present short note should be
considered as a supplement with a focus an the role of meson physics
in this game. The hadronic vacuum polarization (HVP) part I have
reviewed not long ago
in~\cite{Jegerlehner:2015stw,Jegerlehner:2017lbd} and I will be short
on that and focus more on the HLbL part.

\section{To be improved: leading hadronic=mesonic effects}

The problem is a reliable and precise evaluation of the
non-perturbative strong interaction effects. Besides the dispersion
relation (DR) approach applicable where the relevant experimental
cross sections are available one needs low energy effective hadronic
modeling like vector meson dominance (VMD), scalar QED (sQED),
extended Nambu-Jona-Lasinio (ENJL) or hidden local symmetry (HLS) or
similar Resonance Lagrangian Approach models, which attempt to extend
chiral perturbation theory (CHPT) by including vector mesons (VMD) in
accord with the chiral structure of QCD. Lattice QCD ab initio
calculations come closer in precision and already have provided
important constraints and information (see e.g.~\cite{Meyer:2018til}).

The difficulty of getting precise estimate of the non-perturbative
effects I illustrate for the HVP contribution (see Fig.~\ref{fig-01}) 
in the following Table~\ref{tab-1}, with
entries from DR, VMD, sQED and perturbative QCD (pQCD) adopting
alternatively constituent and current quark masses. Only the VMD yields
a reasonable agreement with the data-driven DR method while other estimates 
widely differ and badly fail. This kinds of problems become even more
severe in estimating the HLbL contribution which is a 3 scale problem,
while the HVP is a comparatively simple 1 scale problem.
\begin{table}
\centering
\caption{\small Low energy effective estimates of the leading vacuum
polarization effects $ a^{\rm had}_\mu(\mathrm{vap})$.
For comparison:  $ 5.8420
\power{-8}$ for $ \mu$--loop, $ 5.9041 \power{-6}$ for
$ e$--loop}. 
\label{tab-1}
\begin{tabular}{ccccc}
\hline\noalign{\smallskip}
data+DR & $\rho^0$--exchange & $\pi^\pm$--loop
&\multicolumn{2}{c}{QCD [$u,d$]
quark-loops}\\ {\small [280,810] MeV}& BW+PDG & sQED &
{\small constituent quarks} & {\small current quarks}\\
\noalign{\smallskip}\hline\noalign{\smallskip}
$\mathbf{4.2666 \power{-8}}$&$4.2099 \power{-8}$ & $1.4154 \power{-8}$ &
$2.2511\power{-8}$&$4.4925\power{-6}$ \\
\noalign{\smallskip}\hline
\end{tabular}
\end{table}
\subsection{Leading order HVP}
Adopting the data-driven DR approach the leading hadronic contribution
HPV from the photon vacuum polarization is dominated by the $e^+e^-
\to \pi^+\pi^-$ channel to about 75\%. The major part is determined by the low lying $\rho,\,\omega$
and $\phi$-resonances and in the 1 to 2 GeV region by exclusive
channel data as listed in Table~\ref{tab-2}.  Besides a tiny
contribution from nucleon pair production all kinds of mesonic states
contribute. These have been measured quite exhaustively by
BaBar. Because of the high precision required also small contributions
are to be kept under control.  
{\scriptsize
\begin{table}
\caption{\small Exclusive channels in the range [0.305,1.8] GeV based
on locally weighted averaged data, which compares to a similar
Table~5.3 of~\cite{Jegerlehner:2017gek} where $\omega$ and $\phi$ are
taken as BW resonances using PDG parameters and $\pi\pi$ data from
different experiments are combined by taking weighted averages of
integrals in overlapping regions. The HLS effective theory allows us
to predict the cross sections: \mbo{\ppm,~
\pi^0\gamma,~\eta\gamma,~\eta'\gamma,~\pi^0\pi^+\pi^-,~K^+K^-,~K^0\bar{K}^0}. 
The HLS missing part
\mbo{4\pi,5\pi,6\pi,\eta\pi\pi,\omega\pi} etc., in any case is evaluated using data
directly. The CHPT low energy tail and the final state radiation (FSR)
channel $\pi^+\pi^-\gamma$ are added separately. Values in units $10^{-10}$.}
\label{tab-2}
\centering
\begin{tabular}{lrrr|lrrr}
\hline\noalign{\smallskip}
final state & contrib. & stat & syst & final state & contrib. & stat & syst\\
\hline\noalign{\smallskip}
 $  \pi^0\gamma                   $ &    5.34  &    0.78  &    0.85 &  $  K_SK^\pm\pi^\mp               $ &    0.78  &    0.11  &    0.12 \\
 $  \pi^+\pi^-                    $ &  501.55  &   73.01  &   79.58 &  $  K_SK_L\eta                    $ &    0.12  &    0.02  &    0.02 \\
 $  \pi^+\pi^-\pi^0               $ &   49.15  &    7.15  &    7.80 &  $  K^+K^-\eta                    $ &    0.00  &    0.00  &    0.00 \\
 $  \eta\gamma                    $ &    0.56  &    0.08  &    0.09 &  $  K^+K^-\pi^+\pi^-              $ &    0.32  &    0.05  &    0.05 \\
 $  \pi^+\pi^-2\pi^0              $ &   17.87  &    2.60  &    2.84 &  $  K^+K^-\pi^0\pi^0              $ &    0.04  &    0.01  &    0.01 \\
 $  2\pi^+2\pi^-                  $ &   13.54  &    1.97  &    2.15 &  $  K_SK_L\pi^0\pi^0              $ &    0.11  &    0.02  &    0.02 \\
 $  \pi^+\pi^-3\pi^0              $ &    0.74  &    0.11  &    0.12 &  $  K_SK_L\pi^+\pi^-              $ &    0.08  &    0.01  &    0.01 \\
 $  2\pi^+2\pi^-\pi^0             $ &    0.98  &    0.14  &    0.15 &  $  K_SK_S\pi^+\pi^-              $ &    0.01  &    0.00  &    0.00 \\
 $  2\pi^+2\pi^-\eta              $ &    0.03  &    0.00  &    0.00 &  $  K^+K^-\pi^+\pi^-\pi^0         $ &    0.02  &    0.00  &    0.00 \\
 $  2\pi^+2\pi^-2\pi^0            $ &    0.64  &    0.09  &    0.10 &  $  K_SK^\pm\pi^\mp\pi^0          $ &    0.12  &    0.02  &    0.02 \\
 $  3\pi^+3\pi^-                  $ &    0.11  &    0.02  &    0.02 &  $   \phi \eta                    $ &    0.01  &    0.00  &    0.00 \\
 $  \pi^+\pi^-4\pi^0              $ &    0.03  &    0.00  &    0.00 &  $  \eta \pi^+\pi^-\pi^0          $ &    0.01  &    0.00  &    0.00 \\
 $  \omega\pi^0                   $ &    0.83  &    0.12  &    0.13 &  $   \omega \eta                  $ &    0.02  &    0.00  &    0.00 \\
 $  K^+K^-                        $ &   22.26  &    3.24  &    3.53 &
 \multicolumn{4}{c}{----------------------------}\\
 $  K^0_SK^0_L                    $ &   14.13  &    2.06  &    2.24 &     sum  no CHPT, no FSR          &  630.26    &   91.74  &  100.00 \\
 $  \omega\pi^+\pi^-              $ &    0.01  &    0.00  &    0.00 &     sum                           &  637.12    &   92.74  &  100.00 \\
 $  \eta\pi^+\pi^-                $ &    0.01  &    0.00  &    0.00 &     sum HLS                       &  599.85    &    0.50  &   91.74 \\
 $  K^+K^-\pi^0                   $ &    0.16  &    0.02  &    0.03 &     sum non HLS                   &   37.27    &    2.51  &    5.43 \\
 $  K_SK_L\pi^0                   $ &    0.69  &    0.10  &    0.11 & \multicolumn{4}{c}{}\\ 
\noalign{\smallskip}\hline
\end{tabular}
\end{table}
} Narrow resonances I usually include as Breit-Wigner (BW) states
using PDG parameters. For the low energy region below $1.05~\gv$
(covering $\rho,\,\omega$ and $\phi$) I obtain $a_{\mu}^{\rm
had}[E<1.05~\gv] =505.75_\rho (0.83) (2.57)[2.70]
+ 35.23_\omega (0.29) (0.69)[0.75] + 34.31_\phi (0.27) (0.38)[0.47]=575.29(0.92)(2.92)[2.84]$, 
while using data directly I find $a_{\mu}^{\rm
had}[E<1.05~\gv]=577.46_{\rm data} (0.71) (4.02)[4.08]$ in units
$10^{-10}$ with statistical, systematic and total errors. This
illustrates the fair agreement between different treatments of the
data. The total leading order HVP contribution in the first case
yields $\amuh=688.65 (1.03) (3.82)[3.96] \power{-10}$ while using local averaging of the
$\pi\pi$ data the second yields a slightly higher but less precise $\amuh=690.82 (0.85)
(4.85)[4.92]\power{-10}$. The low energy two body channel together with the $3\pi$
one can be subjected to a global HLS fit~\cite{Benayoun:2012wc} which
yields 83.4\% of the total and as a best fit estimate one finds:
$a_\mu^{\rm had}=(681.9\pm3.2)\power{-10}=(569.04[1.08]_{\rm
HLS-fit}+112.82[3.01]_{\rm HLS-missing}) \power{-10}$. For a comparison with other
results see Fig.~6 in~\cite{Jegerlehner:2018zrj} (see
also~\cite{Ananthanarayan:2013zua,Davier:2017zfy,Keshavarzi:2018mgv}).
    
\subsection{HLbL}
\begin{figure}
\centering
\includegraphics[width=0.48\textwidth]{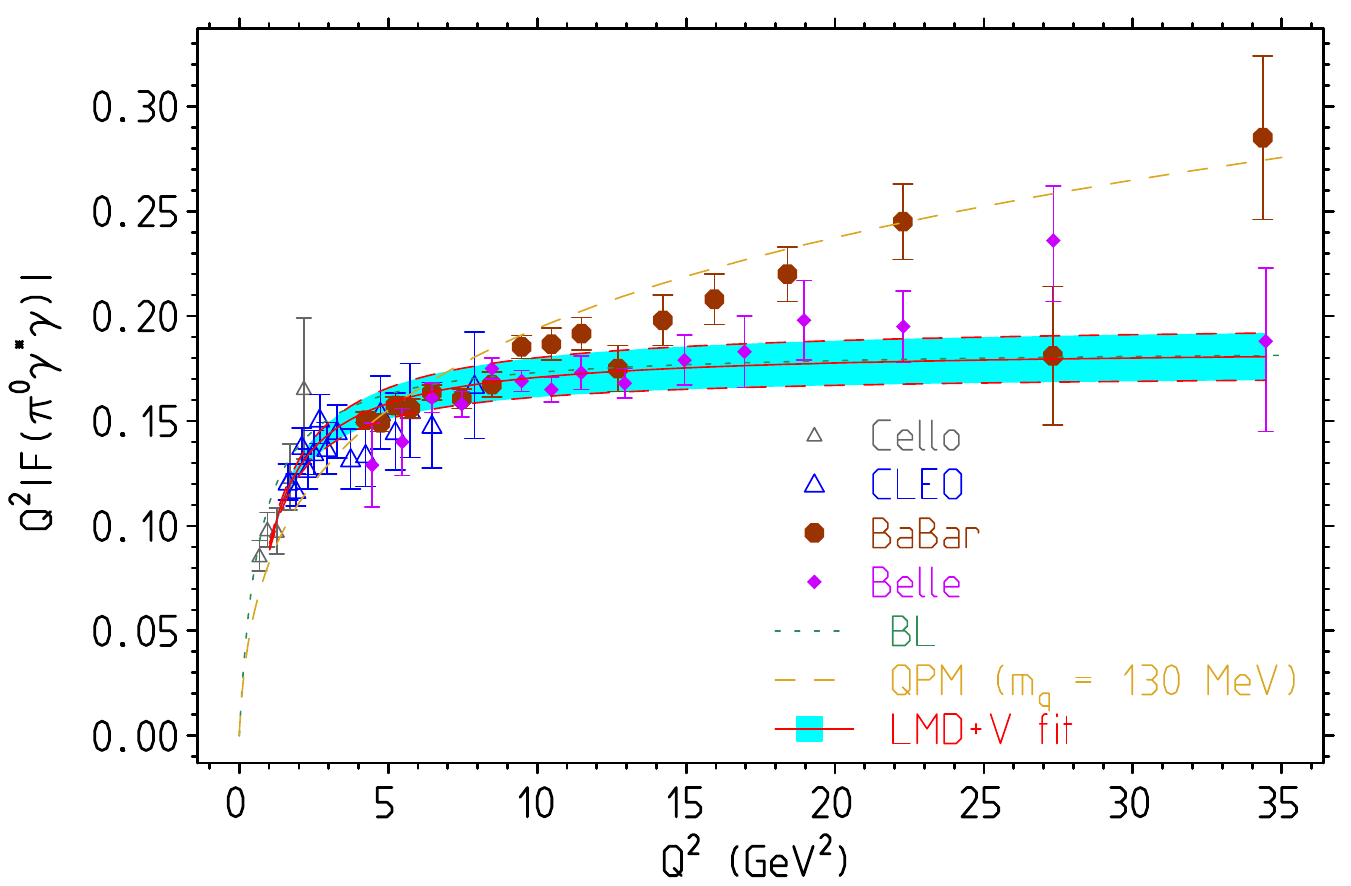}
\includegraphics[width=0.48\textwidth]{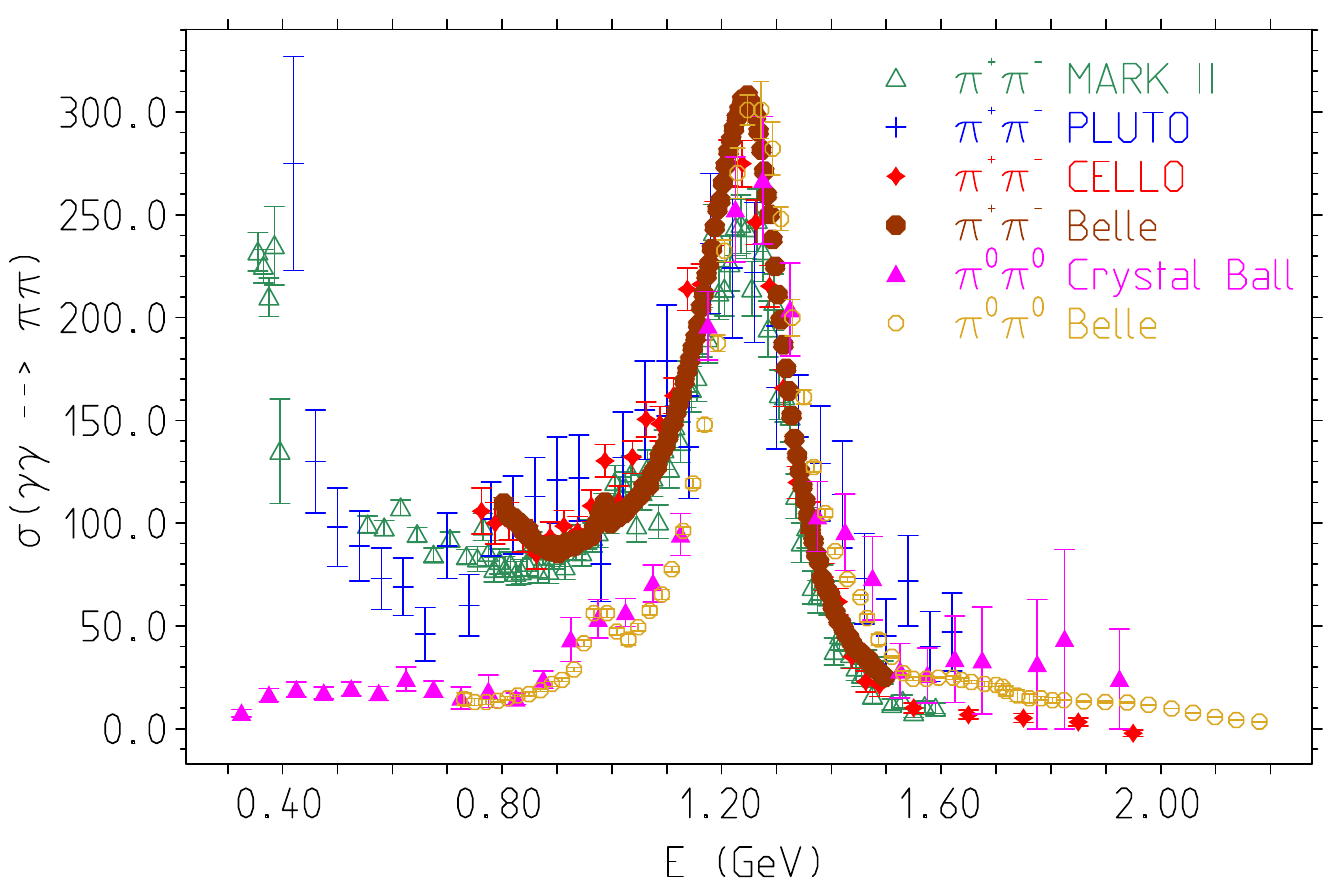}
\caption{\small Left: pion production in $ \gamma\gamma$ 
by CELLO, CLEO, BaBar and Belle measurements of the $\pi^0$ form
factor $\FFac(m_\pi^2,-Q^2,0)$ at high space--like $Q^2$. Towards
higher energies BaBar is somewhat conflicting with Belle. The latter
conforms with theory expectations, which we use as an OPE
constraint. More data are available for $\eta$ and $\eta'$
production. Right: di-pion production in $ \gamma\gamma$ fusion. At
low energy we have direct $ \pi^+\pi^-$ production and by strong
rescattering $ \pi^+\pi^- \to \pi^0\pi^0$, however with very much
suppressed rate. With increasing energy, above about 1 GeV, the strong
$ q\bar{q}$ resonance $ f_2(1270$ appears produced equally at expected
isospin ratio $ \sigma(\pi^0\pi^0)/\sigma(\pi^+\pi^-)=\frac12$.  This
demonstrates convincingly that we may safely work with point-like
pions below 1 GeV.}
\label{fig:pi0ffexp}
\end{figure}

The HLbL contribution is dominated by single particle
exchanges. Thereby $e^+e^- \to e^+e^- \gamma \gamma^* \to
e^+e^-\,\mathrm{\bf hadrons}$ data provide important experimental
constraints on hadronic transition form factors (TFF). As indicated
one of the photons is quasi real in order to get the required
sufficient statistics, while the second is
off-shell. Fig.~\ref{fig:pi0ffexp}-left shows the available data which
constrain the $\pi^0\gamma\gamma^*$ form factor of Fig.~\ref{fig-03}a
and Fig.~\ref{fig:pi0ffexp}-right the pion-loop amplitude of Fig.~\ref{fig-03}b.
Actually, besides the pseudoscalars also axial-, scalar- and
tensor-mesons contribute. An overview of various HLbL one-particle
exchange contributions is given in Table~\ref{tab-3} (see
also~\cite{JN,Nyffeler:2017ohp,Bijnens:2017trn,Knecht:2018sci}). While
the $\pi^0$ exchange contribution clearly dominates, it is obvious
that the other contributions sum to about one-third of the leading one
and have to be determined with comparable precision. This is a highly
non-trivial task and has been estimated by a very few groups (HKS~\cite{HKS95},
BPP~\cite{BijnensLBL,Bijnens:2017trn}, MV~\cite{MV03}) only. The simplest 
channel is the dominant $\pi^0$ one and
has been evaluated by many groups in many different models/approaches
as listed in Table~5.13 of~\cite{Jegerlehner:2017gek} where also the
references are given. The relative stability of the
results is not very surprising because the relevant
$\pi^0\gamma\gamma$ transition form factor is constrained by the known
$\pi^0\to \gamma \gamma$ decay rate, fixing $\FFabc(m_\pi^2,0,0)$, and
by QCD asymptotic behavior of $\FFab(m_\pi^2,0,-Q^2) $ when $Q^2$ gets large,
essentially the Brodsky-Lepage (BL) constraint $\propto 1/Q^2$ as
supported by experimental data (see Fig.~\ref{fig:pi0ffexp}-left). A
new important constraint
has been obtained from lattice QCD by calculating
$\FFa(m_\pi^2,-Q^2,-Q^2)$~\cite{Gerardin:2016cqj}. A first dispersive
calculation of the pion-loop contribution $a_\mu^{\pi-{\rm
box}}+a_{\mu,J=0}^{\pi\pi,\pi-{\rm pole}\,{\rm
LHC}}=-24(1)\power{-11}$ has been presented
in~\cite{Colangelo:2015ama}. A new solid data-driven evaluation of the
pion-pole contribution (in the pion pole approximation) based on the
dispersive model yields a fairly precise $a_\mu^{{\rm
HLbL}\,\pi^0}=6.26^{+0.30}_{-0.25}
\power{-10}$~\cite{Hoferichter:2018kwz}. Also a new estimate of the scalar
contribution $a_\mu^{\rm HLbL-scalars}\simeq (-0.8\pm 7.1)\power{-11}$
has been worked out in~\cite{Knecht:2018sci}. The scalar contribution
should be negative in any case. For more details I
refer to~\cite{Jegerlehner:2018zrj} or to my
book~\cite{Jegerlehner:2017gek}.
\begin{table}[h]
\centering
\caption{Estimates of one-particle exchange contributions to HLbL.
Leading is the $\pi^0$ exchange which amounts to about 65\% of the
total HLbL. Numbers in units $10^{-11}$.}
\label{tab-3}
\begin{tabular}{llrrrr}
\hline
\multicolumn{2}{c}{contribution} & \multicolumn{3}{c}{individual} & \multicolumn{1}{c}{total} \\
\hline
pseudoscalars &
\mbo{a_{\mu}^{\mbox{\tiny{LbL}}}(\pi^0,\eta,\eta')}
& \textbf{64.68}&14.87&15.90 & \mbo{95.45\pm 12.40\py}\\
axials &
\mbo{a_{\mu}^{\mbox{\tiny{LbL}}}(a_1,f_1,f_1')}& 1.89\px&5.19&0.47 &\mbo{7.55\pm\px2.71\py}\\
scalars &
\mbo{a_{\mu}^{\mbox{\tiny{LbL}}}(a_0,f_0,f_0')}&
-0.17\px&-2.96&-2.85&\mbo{-5.98\pm \px1.20\py}\\
tensors &
\mbo{a_\mu^{\mbox{\tiny{LbL}}}(f_2',f_2,a_2')} & 0.79\px & 0.07 & 0.24
&\mbo{1.1\px \pm \px 0.1\px \py}\\
\multicolumn{2}{c}{sum single meson exchange}&&&&\mbo{98.12\pm 12.75\py}\\
\multicolumn{2}{c}{+ $\pi^\pm,K^\pm$ loops + quark loops}&&&&\mbo{103.40\pm 28.80\py}\\
\hline
\end{tabular}
\end{table}
 My estimate is $ a_\mu^{\rm HLbL}=[{ 95.45(12.40)}{+7.55(2.71)}{ -5.98(1.20)}{
+20(5)}{-20(4)}{+2.3(0.2)}{ +1.1(0.1)}{ +3(2)}]\power{-11}
=103.4(28.8)\power{-11}$. For a comparison with other evaluations see
Table~5.19 and Fig.~5.66 of my book~\cite{Jegerlehner:2017gek}.
Agreement between different estimates is not yet satisfactory, and a
reduction of the errors is still a major issue. Progress we expect
from lattice QCD and/or from the dispersive approach (Colangelo et
al.~\cite{Colangelo:2015ama}, Pauk and
Vanderhaeghen~\cite{Pauk:2014rta}), which is determining the various
HLbL amplitudes based on data and DR's.
\section{Summary and conclusion}
The relevance of different mesonic effects 
in relation to the new experimental result to come
are tabulated in Table~\ref{tab-4}.
\begin{table}
\centering
\caption{Significance of typical channels at present and as expected
after the new experimental result. In units $10^{-10}$
or in Standard Deviations (SD), {\em present} relative to present combined
uncertainty $7.6\power{-10}$, {\em coming} relative to expected experimental
uncertainty $1.6\power{-10}$. The table (bold entries) shows where uncertainties have
to be reduced.}
\label{tab-4}
\begin{tabular}{lcrrr}
\hline
\multicolumn{2}{c}{type} & contribution & SD present & SD coming\\ 
\hline
HVP & { LO} {$O(\alpha^2)$} & 689.5[3.3] & 90.7[0.4]  & 431[{\bf 2.1}] \\
    & \mbo{\pi^+\pi^-} & 505.7[2.7] & 66.6[0.4] & 316.1[{\bf 1.7}] \\
    & \mbo{K^+K^-} & 22.0[0.7] & 2.9[0.1] & 13.8[0.4] \\
    & \mbo{\pi^+\pi^-2\pi^0} & 20.4[0.9] & 2.6[0.1] & 12.8[0.6] \\
    & { 1.05-2\gv} & 62.2[2.5] & 8.2[0.3]& 38.9[{\bf 1.6}]\\
    & { HO} {$O(\alpha^n)$} ($n>2$) & -8.7[0.1] & 1.1[0.0] & 5.4[0.0] \\ 
HEW & { 3 families} & -1.5[0.0] & \multicolumn{2}{c}{small by anomaly cancellation}\\
HLbL& { all} {$O(\alpha^3)$} & 10.3[2.9] &1.4[0.4] & 6.4[{\bf 1.2}] \\
    & \mbo{\pi^0} & 6.3[0.8] & 0.9[0.1] & 4.1[0.5]\\  
\hline
\end{tabular}
\end{table}
The present status of the SM prediction of $a_\mu$ is summarized in Table~\ref{tab:amuSM}.
{\small
\begin{table}[h]
\centering
\caption{Standard model theory and experiment
comparison [in units $10^{-10}$].}
\label{tab:amuSM}
\begin{tabular}{lr@{.}lr@{.}lc}
\hline
Contribution & \multicolumn{2}{c}{Value$\times 10^{10}$} & \multicolumn{2}{l}{Error$\times 10^{10}$} & Reference \\
\hline
QED incl. 4-loops + 5-loops & 11\,658\,471&886 & 0&003 &
\cite{Aoyama12am,Laporta:2017okg}  \\
Hadronic LO vacuum polarization & 689&46 &  3&25 & \cite{Jegerlehner:2017zsb} \\
Hadronic light--by--light &   10&34 & 2&88 & \cite{Jegerlehner:2017gek}\\
Hadronic HO vacuum polarization & -8&70 & 0&06 & \cite{Jegerlehner:2017zsb} \\
Weak to 2-loops & 15&36 & 0&11 & \cite{Gnendiger:2013pva}
  \\
\hline
 Theory & \multicolumn{2}{l}{11\,659\,178.3} & \multicolumn{2}{l}{4.3} & --  \\
Experiment & 11\,659\,209&1 & 6&3 & \cite{Bennett:2006fi}  \\
The. - Exp.   {{4.0}} standard deviations &-30&6 & 7&6 & -- \\ \hline
\end{tabular}
\end{table}}
What the 4 $\sigma$ deviation is about? Is it  new physics?  a statistical fluctuation?
underestimating uncertainties (experimental, theoretical)?  do experiments measure what
theoreticians calculate? Is the Bargmann-Michel-Telegdi equation of a Dirac particle in external
electromagnetic fields sufficiently accurate? Could real photon
radiation affect the measurement?

A ``New Physics'' interpretation of the persisting 3 to 4 $\sigma$
deviation requires relatively strongly coupled states in the range
below about 250 GeV. The problem is that LEP, Tevatron and LHC direct
bounds on masses of possible new states $X$ typically say \mbo{M_X>
800\gv}. In any case $\amu$ constrains BSM scenarios distinctively
and at the same time challenges a better understanding of the SM
prediction. 

Progress on the theory side requires more/better data and/or progress
in non-perturbative QCD. The muon \mbo{g-2} prediction is limited by
hadronic uncertainties, which are dominated by meson form
factors uncertainties. Substantial progress would be possible if one could reach
better agreement on what QCD predicts for the various meson form
factors. Most important is the pion sector be it the $\gamma
\pi^+\pi^-$ or the $\gamma \gamma \pi^0, \gamma \gamma
\pi^+\pi^-, \gamma \gamma \pi^0 \pi^0$ and related TFFs. 
A big challenge for the meson physics community. The most promising
dispersive methods require primarily improved data, which is not easy
to get.

Fortunately, lattice QCD is making big progress and begins to help to
settle hadronic issues.  For both of the critical contributions HVP
and HLbL lattice QCD will be the answer one day
(see~\cite{Meyer:2018til} and references therein), I expect.  But a
lot remains to be done while a new $a^{\rm exp}_\mu$ is on the way!

\noindent
{\bf Acknowledgments} \\ Many thanks to the organizers for the kind
invitation to the MESON 2018 Conference and for giving me the opportunity to present
this talk and thank you so much for your kind support.

\end{document}